\documentclass[preprint,showpacs,aps]{revtex4}
\usepackage{graphicx}% Include figure files
\def\RR{\hbox{{\rm I}\kern-.2em\hbox{\rm R}}}
\def\pRR{\hbox{{\tiny \rm I}\kern-.1em\hbox{{\tiny \rm R}}}}

\def\NN{\hbox{I\kern-.2em\hbox{N}}}

\begin{document}

\title{Three eras of micellization}
\renewcommand{\thefigure}{\arabic{section}.\arabic{figure}}
\renewcommand{\theequation}{\arabic{section}.\arabic{equation}}
\author{ J. C. Neu\cite{neu:email}}
\affiliation{Department of Mathematics, University of California
at Berkeley, Berkeley, CA 94720, USA}
\author{ J. A. Ca\~nizo\cite{canizo:email}}
\affiliation{Departamento de Matem\'atica Aplicada, Universidad
de Granada, E-18071 Granada, Spain}
\author{ L. L. Bonilla\cite{bonilla:email}}
\affiliation{Departamento de Matem\'aticas, Universidad Carlos
III de Madrid, Avda.\ Universidad 30, E-28911 Legan{\'e}s, Spain}
\date{August 24, 2002}

\begin{abstract} 
Micellization is the precipitation of lipids
from aqueous solution into aggregates with a broad distribution
of aggregation number. Three eras of micellization are
characterized in a simple kinetic model of Becker-D\"oring type.
The model asigns the same constant energy to the $(k-1)$
monomer-monomer bonds in a linear chain of $k$ particles. The
number of monomers decreases sharply and many clusters of small
size are produced during the first era. During the second era,
nucleii are increasing steadily in size until their distribution
becomes a self-similar solution of the diffusion equation.
Lastly, when the average size of the nucleii becomes comparable
to its equilibrium value, a simple mean-field Fokker-Planck
equation describes the final era until the equilibrium
distribution is reached.

\end{abstract}

\pacs{82.70.Uv, 83.80.Qr, 05.40.-a, 05.20.Dd}
\maketitle

\setcounter{equation}{0}
\setcounter{figure}{0}
\section{Introduction}
\label{sec-introduction}

Spontaneous self-assembly of small molecular aggregates in
aqueous solutions forms association colloids or complex fluids
\cite{isr91}. Depending on their mean aggregation number,
molecular volume and critical hydrocarbon chain length, lipids
can pack into spherical or cylindrical micelles. The surfaces of
these structures are formed by the hydrophilic heads of the
monomer molecules, whose hydrophobic tails lie inside the
aggregate. Equilibrium thermodynamics shows that rod-like
cylindrical aggregates have a polydisperse distribution of sizes
(micellization), whereas the sizes of spherical aggregates grow
indefinitely (phase segregation) \cite{isr91}. The latter process
is similar to other examples of first order phase transitions
\cite{ll10} such as condensation of liquid droplets from a
supersaturated vapor, colloidal crystallization \cite{gas01} and
the segregation by coarsening of binary alloys quenched into the
miscibility gap \cite{LS,XH,mar96}. Understanding the kinetics of
nucleation and growth beyond the determination of the
steady-state nucleation rate is a task of great importance and
not yet completely accomplished. This is so despite the large
literature on nucleation and growth \cite{kel91}, and several
attempts at bridging the gap between nucleation and late-stage
coarsening theories \cite{juanjo}.

In this paper, we study asymptotically a simple discrete model of
micellization kinetics of Becker-D\"oring type
\cite{kel91,juanjo,pen83,gan99}. Starting from an initial
condition of pure monomers, we expect the system to evolve to
the well known polydisperse equilibrium distribution
\cite{isr91}. However, the nonequilibrium evolution is
interesting per se and because the methodology employed here may
be applicable to the kinetics of phase segregation. We find that
the approach to equilibrium occurs in three well defined stages
or eras. Starting from the initial state of pure monomers, the
number of monomers decreases sharply and many clusters of small
size are produced during the first era. During the second era,
aggregates are increasing steadily in size until their
distribution becomes a self-similar solution of the diffusion
equation. Lastly, when the average size of the nucleii becomes
comparable to its equilibrium value, a simple mean-field
Fokker-Planck equation describes the final era until the
equilibrium distribution is reached. Numerical solution of the
model confirms all the theoretical predictions.

The rest of the paper is as follows. In Section \ref{sec:model},
we review the equilibrium properties of self-assembling
aggregates and introduce discrete kinetic models of
Becker-D\"oring type to describe them. Depending on the binding
energy of the aggregate with $k$ monomers ($k$ cluster),
micellization or phase segregation occurs. For rod-like
aggregates, the binding energy of a $k$ cluster (relative to
isolated monomers in solution) is $(k-1)$ times the
monomer-monomer bond energy, and an equilibrium size
distribution exists (micellization). For spherical aggregates,
the binding energy includes a term proportional to the surface
area of the aggregate and no equilibrium size distribution
exists beyond a critical density. Then aggregates grow
indefinitely and phase segregation occurs following the typical
nucleation and growth kinetics. Section \ref{sec:numerics}
presents a numerical simulation of micellization kinetics which
clearly revels its three eras. The agenda of the asymptotic
analysis is now clear, and is carried out in Section
\ref{sec:asymptotics}. The last section contains our conclusions
and suggestions for experiments.

\setcounter{equation}{0}
\setcounter{figure}{0}
\section{Thermodynamics and kinetic models}
\label{sec:model} 
The model presented here is nucleation in a
lattice. There are systems, such as proteins aggregating in a
cubic phase of lipid bilayers, for which a lattice formulation
is physically correct. In this paper, the main reasons for a
lattice model are clarity, and the expectation that the dilute
limit of the lattice model (in which there are many more binding
sites, $M$, than particles, $N$) should closely resemble
crystallization from a dilute solution. The latter is a
classical problem in the kinetic theory of first order phase
transitions \cite{ll10}. We shall now review the equilibrium
statistical mechanics of aggregates, distinguishing between
micellization and phase segregation, and then introduce the
kinetic models we study.

\subsection{Equilibrium size distribution of aggregates}  
Let us assume that we have $p_k\geq 0$ clusters with $k$ particles
(in short, $k$ clusters) so that
\begin{eqnarray} 
N=\sum_{k=1}^N k p_k. \label{e1}
\end{eqnarray} 
Let $e_k$ be the energy of a $k$ cluster. The total energy of the
lattice system is
\begin{eqnarray} E = \sum_{k=1}^{N} p_k e_k = N\, e_1 +
\sum_{k=2}^{N} p_k (e_k - k e_1), \label{e2}
\end{eqnarray} 
where we have used the particle conservation (\ref{e1}). Except
for a constant $Ne_1$, the total energy is
\begin{eqnarray} 
E = - \sum_{k=2}^{N} p_k \varepsilon_k , \label{e3}\\
\varepsilon_k = k\, e_1 - e_k.\label{e4}
\end{eqnarray} 
Now $E$ is the total lattice energy measured with
respect to a configuration in which all clusters are monomers,
and $\varepsilon_k$ is the binding energy of the $k$ cluster
(notice the sign convention). We will obtain the equilibrium
configuration by minimizing the free energy density with respect
to the density of $k$ clusters. To calculate the entropy, we
proceed as follows. Let $n_j\geq 0$ be the occupation number of
the site $j$, $j=1, \ldots, M$. The configuration space of the
lattice consists of all $M$-tuples of occupation numbers,
$\{n_1,\ldots, n_M\}$, with $\sum_{j=1}^M n_j = N$ and $N\ll M$.
Clearly, there are many {\em indistinguishable} configurations
that produce the same given set of numbers, $p_1,\ldots, p_N$.
Their number $\Omega$ is given by the Bose-Einstein counting
argument,
\begin{eqnarray}
\Omega = {M!\over p_{1}!\ldots p_{N}!\, (M-p_{1}-\ldots -p_{N})!}
, \label{e5}
\end{eqnarray} and the entropy of the system is $k_B \ln\Omega$.
In the appropriate thermodynamic limit, $N\to\infty$ with fixed
densities
$\rho\equiv N/M$ (particles) and $\rho_k\equiv p_k/M$ ($k$
clusters), particle conservation becomes
\begin{eqnarray}
\sum_{k=1}^\infty k \rho_k = \rho , \label{e6}
\end{eqnarray} and  we can show that the entropy density is
\begin{eqnarray} 
{\cal S} &\equiv & {k_{B}\over M}\, \ln\Omega \sim - k_{B}\left(
\sum_{k=1}^\infty \rho_k \ln\rho_k + r\ln r\right), \label{e7}\\
r & = & 1- \sum_{k=1}^\infty \rho_k ,  \label{e8}
\end{eqnarray} 
by using Stirling's formula. The free energy density, $f= E/M-T
{\cal S}$, can be written in terms of $\rho$ and the densities of
clusters having two or more particles by using its definition and
Equations (\ref{e3}) and (\ref{e6}) to (\ref{e8}). The result is
\begin{eqnarray} f &=& - \sum_{k=2}^\infty \rho_k \varepsilon_k
+ k_BT\sum_{k=1}^\infty \rho_k \ln\rho_k + k_BT r\ln r ,
\label{e9}
\end{eqnarray}
where $\rho_1 = \rho - \sum_{k=2}^\infty k \rho_k$ and $r= 1-
\sum_{k=1}^\infty \rho_k$. In the dilute limit, $1-r=
\sum_{k=1}^\infty \rho_k< \sum_{k=1}^\infty k \rho_k = \rho \ll
1$, and therefore $r\sim 1$, $r\ln r \sim -\sum_{k=1}^\infty
\rho_k$, and Eq.\ (\ref{e9}) becomes
\begin{eqnarray}
f &=& - \sum_{k=2}^\infty \rho_k \varepsilon_k + k_BT
\sum_{k=1}^\infty \rho_k\, (\ln\rho_k -1) ,
\label{e10}
\end{eqnarray}
which corresponds to the Boltzmann counting. The equilibrium
density of $k$ clusters ($k\geq 2$) can be found by
differentiating this equation with respect to $\rho_k$ and
equating the result to zero. Taking into consideration that
$\partial \rho_1/\partial\rho_k = -k$ ($k\geq 2$), we obtain
\begin{eqnarray}
\tilde{\rho}_k = \rho_1^k\, \exp\left({\varepsilon_{k}\over
k_{B}T}
\right)   \label{e11}
\end{eqnarray}
(the positive sign in the argument of the exponential is due to
our definition of the binding energies). Eq.\ (\ref{e11}) can be
rewritten as
\begin{eqnarray}
\tilde{\rho}_k &=& \exp\left(-{g_{k}\over k_{B}T}\right) ,
\label{e12}\\ g_k &=& - \varepsilon_k + k_B T k
\ln\left({1\over\rho_1}\right) .  \label{e13}
\end{eqnarray}
$g_k$ as a function of $k$ can be interpreted as the activation
energy of nucleation theory. The equilibrium density of monomers
can be found by inserting Eq.\ (\ref{e11}) into Eq.\ (\ref{e6})
and solving the resulting self-consistent equation for $\rho_1$
in terms of the constant density $\rho$:
\begin{eqnarray}
 \sum_{k=1}^\infty k\, \rho_1^k\, \exp\left({\varepsilon_{k}
\over k_{B}T}\right) = \rho . \label{e14}
\end{eqnarray}
Whether this self-consistent equation has a
solution depends on the value of $\rho$ and on the model we
adopt for the binding energy of a $k$ cluster. Typical models
are as follows. For rod-like aggregates,
\begin{eqnarray}
\varepsilon_{k} = (k-1) \alpha k_B T,  \label{e15}
\end{eqnarray} where $\alpha k_B T$ is the monomer-monomer
bonding energy
\cite{isr91}. For spherical aggregates,
\begin{eqnarray}
\varepsilon_{k} \sim (k-1) \alpha k_B T - {3\over 2}\sigma
k^{{2\over 3}}  \label{e16}
\end{eqnarray}
for $k\gg 1$. Here $\sigma=2\gamma (4\pi v^2/3)^{{1\over 3}}$,
where $\gamma$ and $v=V/M$ are the interfacial free energy per
unit area (surface tension) and the molecular volume,
respectively.

Inserting Eq.\ (\ref{e15}) in Eq.\ (\ref{e14}) and using $\sum_{
k=1}^\infty k x^k = x {d\over dx}\sum_{k=1}^\infty x^k = x/(1-x
)^2$, we obtain
\begin{eqnarray}
\rho = {\rho_{1}\over (1- \rho_1 e^\alpha)^2}.    \label{e17}
\end{eqnarray} This equation has the unique solution
\begin{eqnarray}
\rho_1= {1 + 2\rho e^{\alpha} - \sqrt{1+ 4\rho e^{\alpha}} \over
2\rho e^{2\alpha}},   \label{e18}
\end{eqnarray} with $\rho_1 < e^{-\alpha}$ for all values of the
density $\rho$
\cite{isr91}. Notice that
\begin{eqnarray}
\langle k\rangle \equiv {\sum_{k=1}^\infty k\tilde{\rho}_k\over
\sum_{k=1}^\infty \tilde{\rho}_k} = {\sqrt{1+ 4\rho e^{\alpha}} -
1 \over 2}   \label{e19}
\end{eqnarray}
is the average cluster size in equilibrium.
Notice that, for $\rho e^{\alpha} \gg 1$, $\langle k\rangle \sim
\sqrt{\rho e^{\alpha}}$ and $\tilde{\rho}_k \sim e^{-\alpha}
e^{-k/\langle k \rangle}$.

For spherical aggregates, the self-consistency condition based
on the approximation to $\varepsilon_k$ in Eq.\ (\ref{e16}) is
\begin{eqnarray}
\rho_1\, \sum_{k=1}^\infty k\, \left(\rho_1
e^\alpha\right)^{k-1}\,
\exp\left(-{3\sigma k^{{2\over 3}}\over 2k_B T} \right) = \rho .
\label{e20}
\end{eqnarray}
Clearly this series converges provided $\rho_1
e^\alpha<1$ and it diverges if $\rho_1 e^\alpha > 1$. The
critical monomer concentration $\rho_1 = e^{-\alpha}$ is called
{\em critical micelle concentration} (CMC) \cite{isr91}. Below
CMC eq.\ (\ref{e20}) can be solved for $\rho_1$ and the
aggregates eventually form micelles with an equilibrium size
distribution, whereas phase segregation and indefinite aggregate
growth results if more monomers are added above the CMC. For
$k\gg 1$, the free energy (\ref{e13}) is $g_k \sim \alpha k_B
T + 3\sigma k^{{2\over 3}}/2 - k\varphi$, with $\varphi = k_B T
\ln (\rho_1 e^\alpha)$. For $\varphi >0$, $g_k$ increases for
small
$k$, it has a maximum at the critical cluster size $k_c \approx
(\sigma/\varphi)^3$, and then it decays monotonically as $k$
further increases.

\subsection{Kinetic models} 
Let us now formulate the kinetic theory of aggregation in these
systems. As in the Becker-D\"oring kinetic theory, we shall
assume that a $k$ cluster can grow or decay by capturing or
shedding one monomer at a time. Then
\begin{eqnarray}
\dot{\rho}_{k} = j_{k-1}- j_k\equiv - D_-\, j_k,
\quad k\geq 2,    \label{bd1}\\ j_{k} = d_{k}\,\left\{ e^{{D_+
\varepsilon_{k}\over k_B T}}\, \rho_1 \rho_k -  \rho_{k+1}
\right\},
\label{bd2}
\end{eqnarray}
or finally,
\begin{eqnarray} j_{k} = d_{k}\,\left\{ \left(e^{-{D_+
g_{k}\over k_B T}} - 1\right)\, \rho_k - D_+\rho_{k} \right\},
\label{bd3}
\end{eqnarray}
Here $D_+\varepsilon_k\equiv
\varepsilon_{k+1}-\varepsilon_k = - D_+ g_k + k_B
T\ln(1/\rho_1)$ and $j_k$ is the net rate of creation of a $k+1$
cluster from a $k$ cluster, given by the mass action law. We
have made the detailed balance assumption to relate the kinetic
coefficient for monomer aggregation to that of decay of a
$(k+1)$ cluster, $d_{k}$. Then $\tilde{\rho}_k$ given by Eq.\
(\ref{e11}) solves $j_k =0$. The kinetic model is described by a
closed system of equations once we supplement Eqs.\ (\ref{e6}),
(\ref{bd1}) and (\ref{bd2}) with expressions for the binding
energy of a $k$ cluster, $\varepsilon_k$ and for the kinetic
coefficient of the decay reaction, $d_{k}$.

The simplest possible model for micellization within the
Becker-D\"oring theory is obtained by setting $\varepsilon_k =
(k-1) \alpha k_BT$ and $d_k=1$ in Eq.\ (\ref{bd2}) for the
creation rate of a $(k+1)$ cluster (rescaling of time can absorb
a constant cluster decay rate $d_k= d$; typical time scales
describing aggregation kinetics range from microseconds to
milliseconds \cite{ani74,hun87,sch02}). Equations (\ref{bd1}) and
(\ref{bd2}) then become the following discrete Smoluchowski
equation:
\begin{eqnarray}
\dot{\rho}_{k} +\left(e^\alpha \rho_1 - 1\right)\, (\rho_k -
\rho_{k-1}) = \rho_{k+1} -2\rho_k + \rho_{k-1} ,  \label{bd4}
\end{eqnarray}
to be solved together with the conservation condition (\ref{e6}),
namely, $\sum_{k=1}^\infty k \rho_k = \rho$. At $t=0$, we assume
that $\rho_k= \rho \delta_{k1}$. We shall consider the limit
$\rho \gg e^{-\alpha}$, in which the initial monomer
concentration is much larger than the CMC. The parameters $\rho$
and $\alpha$ are not really independent: if we rescale the
cluster densities with $\rho$, so that
\begin{eqnarray}
\rho_k=\rho r_k, \label{bd5}
\end{eqnarray}
and define a scaled time,
\begin{eqnarray}
\tau \equiv e^\alpha \rho\, t\equiv {t\over\epsilon},
\label{bd8}
\end{eqnarray}
the rescaled problem contains the single parameter
$\epsilon\equiv (\rho e^\alpha)^{-1}\ll 1$. Then Eqs.\
(\ref{bd4}) and (\ref{e6}) become
\begin{eqnarray}
{dr_{k}\over d\tau} &+& (r_1 - \epsilon)\, (r_k - r_{k-1}) =
\epsilon\, (r_{k+1} -2r_k + r_{k-1}) ,\quad k\geq	2 \label{bd9}\\
1  &=& \sum_{k=1}^\infty k\, r_k , \label{bd12}
\end{eqnarray}
to be solved with initial conditions
\begin{eqnarray}
r_1(0)=1,\,\, r_2(0) =r_3(0)=\ldots =0 . \label{bd6}
\end{eqnarray}
Lastly, notice that we can straightforwardly derive two global
identities from Eqs.\ (\ref{bd9}) and (\ref{bd12}):
\begin{eqnarray}
{dr_{1}\over d\tau} & +& r_1 (r_1+ r_c) + \epsilon\, (r_1 - r_2 -
r_c) = 0 ,  \label{bd10}\\
{dr_{c}\over d\tau} &+& r_1 r_c + \epsilon\, (r_1 -r_c) = 0 .
\label{bd11}
\end{eqnarray}
Here $r_c$ is the total density of clusters
\begin{eqnarray}
r_{c}= \sum_{k=1}^\infty\, r_k ,
\label{bd7}
\end{eqnarray}
and, initially, $r_c(0)=1$.

\setcounter{equation}{0}
\setcounter{figure}{0}
\section{Numerical results}
\label{sec:numerics}
\begin{figure}
\begin{center}
\includegraphics[width=7cm]{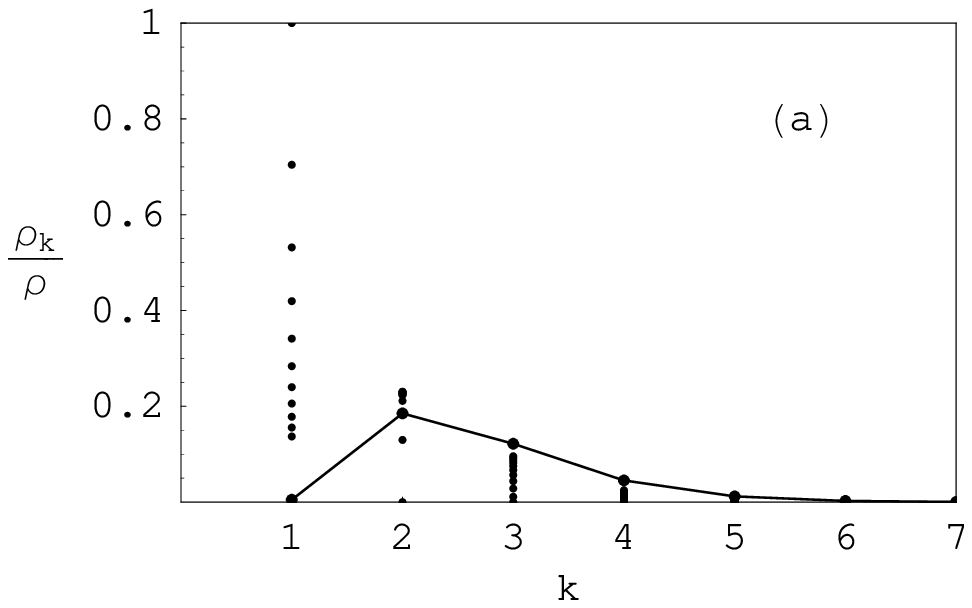}
\includegraphics[width=7cm]{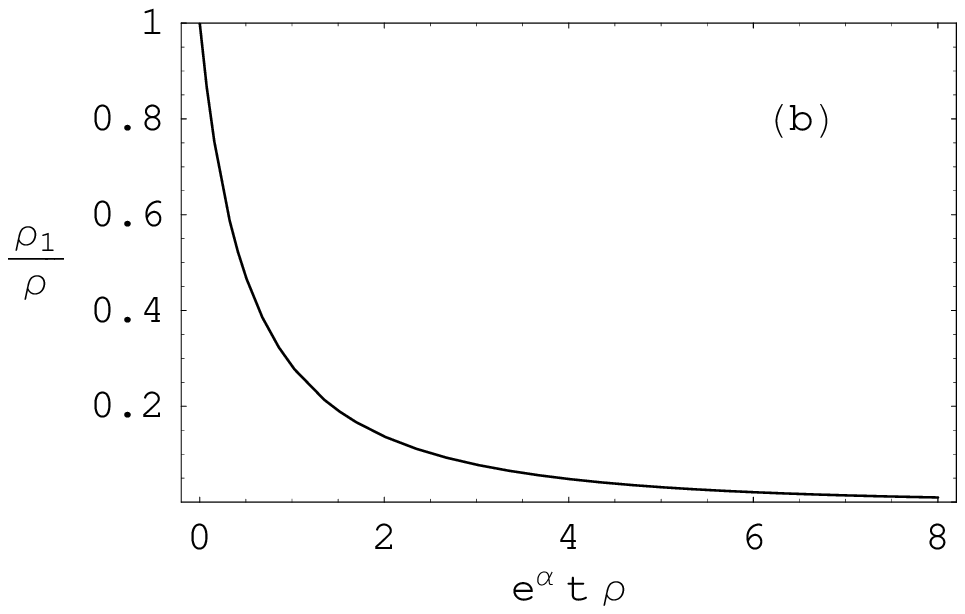}
\includegraphics[width=7cm]{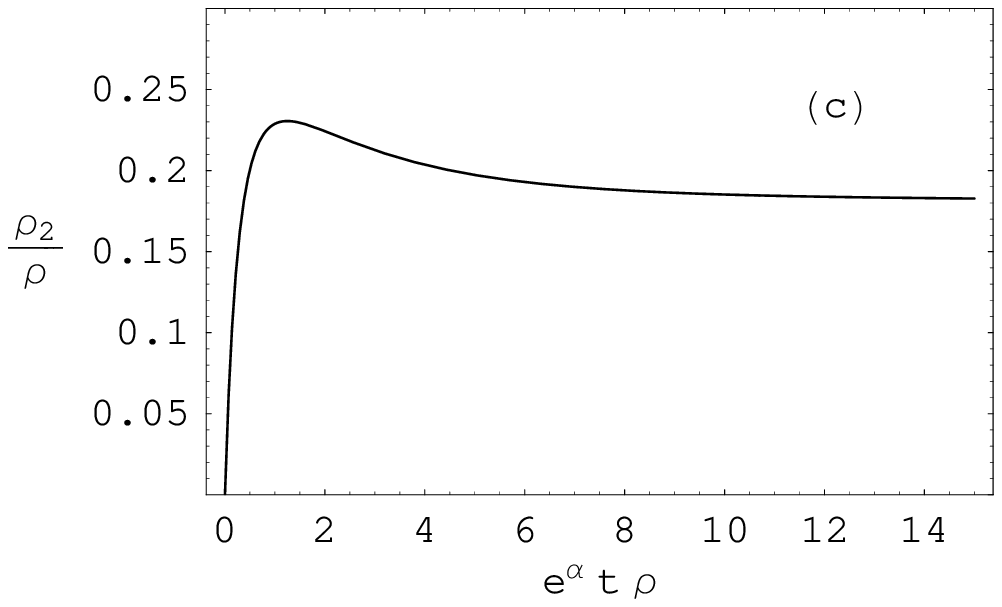}
\vspace{0.5 cm}
\caption{(a) Scaled cluster size distribution $\rho_k/\rho$ as a
function of $k$ for $0\leq\tau\leq 10$. At time $\tau=10$, the
values of $\rho_1/\rho$, $\rho_2/\rho$, etc.\ have been joined by
straight lines as a guide for the eye. (b) Evolution of the scaled
monomer concentration, $\rho_1/\rho$. (c) Evolution of the scaled
dimer concentration, $\rho_2/\rho$. Parameter values are
$\alpha =10$ and $\rho=0.1$.}
\label{fig1}
\end{center}
\end{figure}

\begin{figure}
\begin{center}
\includegraphics[width=7cm]{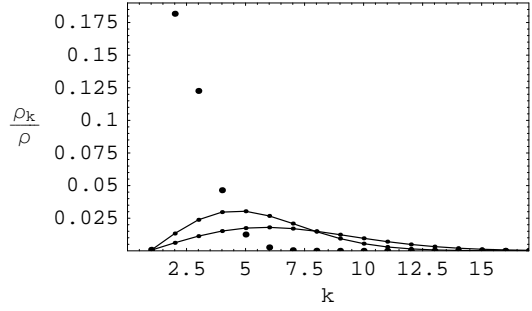}
\vspace{0.5 cm}
\caption{Same as Fig. \ref{fig1}(a), for the times $\tau=20$,
$10^4$ and $2\times 10^4$. At the two later times, we have joined 
values of $\rho_k/\rho$ corresponding to neighboring $k$'s by
straight lines as a guide for the eye. }
\label{fig2}
\end{center}
\end{figure}

\begin{figure}
\begin{center}
\includegraphics[width=7cm]{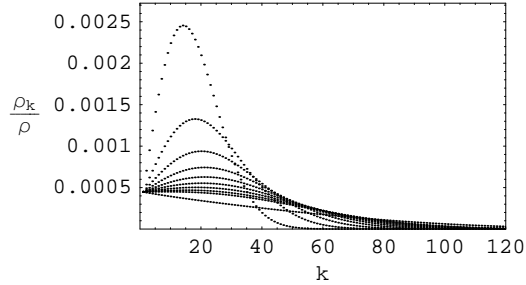}
\vspace{0.5 cm}
\caption{Same as Fig. \ref{fig1}(a), starting at $\tau= 2\times
10^5$. Snapshots of the size distribution have been taken at time
intervals of $\tau= 2\times 10^5$, until a time $\tau= 16\times
10^5$. Then the last snapshot corresponds to $\tau= 40\times
10^5$.}
\label{fig3}
\end{center}
\end{figure}

\begin{figure}
\begin{center}
\includegraphics[width=7cm]{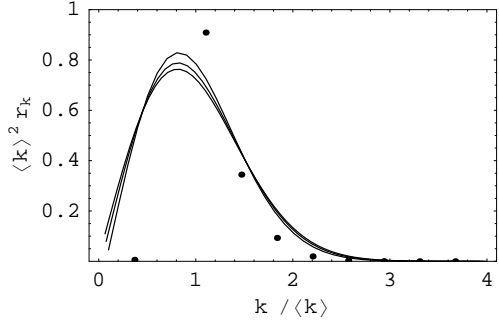}
\vspace{0.5 cm}
\caption{Approximate self-similar behavior of the size
distribution at times $\tau = 50,000$, 100,000 and 150,000 (solid
lines). Notice that $\langle k\rangle^2 r_k$ is approximately the
same function of $k/\langle k\rangle$ at different times. The
dots correspond to $\tau=20$. }
\label{fig5}
\end{center}
\end{figure}

\begin{figure}
\begin{center}
\includegraphics[width=7cm]{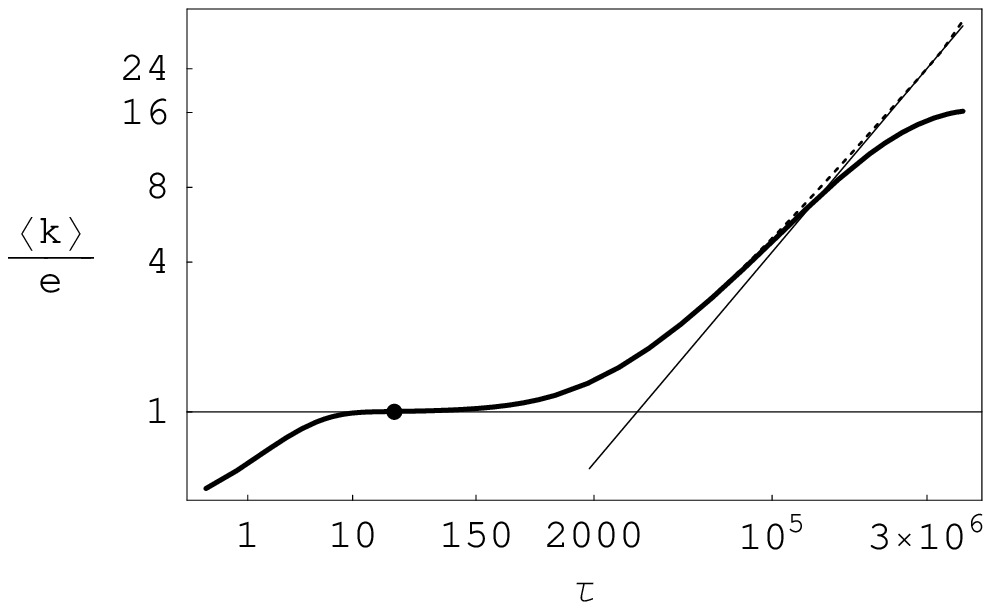} 
\vspace{0.5 cm}
\caption{Evolution of the average cluster size, $\langle k\rangle
/e$, as a function of the scaled time $\tau$ (thick solid line).
The dotted line corresponds to the solution of the system
(\ref{bd23}) in Section \ref{sec:asymptotics} below with an
initial condition corresponding to the dot. The straight line of
slope 1/2 corresponds to the self-similar continuum size
distribution given by Eq.\ (\ref{bd31}). }
\label{fig4}
\end{center}
\end{figure}

Numerical solution of the initial value problem given by Eqs.\
(\ref{bd9}) - (\ref{bd6}) clearly expresses the phenomenology of
micellization, and informs the singular perturbation analysis
carried out in Section \ref{sec:asymptotics}. Figures \ref{fig1}
to \ref{fig4} illustrate the evolution of the size distribution
for $\epsilon=4.54\times 10^{-4}$ (corresponding to $\alpha =10$
and $\rho=0.1$). Figures \ref{fig1}(a), \ref{fig2}, \ref{fig3} and
\ref{fig5} are histograms of $r_k$ as a function of $k$ at
different times, and Figure \ref{fig4} records the time
dependent behavior of the average cluster size, $\langle k
\rangle$. 

Figure \ref{fig1}(a) depicts an early stage of the kinetics. The
sequences of small dots at each $k$ record the values of $r_k$ at
times between $\tau=0$ and $\tau=2$, in increments of
$\Delta\tau= 0.2$, and the larger dots joined by straight lines
record the values of $r_k$ at $\tau=10$. The direction of
increasing time is generally clear: As indicated in Fig.
\ref{fig1}(b), the monomer concentration rapidly decreases to a
small fraction of its initial value $r_1=1$, so that the time
orientation on the line $k=1$ is downward. Many small clusters of
sizes $k$, $2\leq k\leq 5$ are simultaneously created so the time
orientation on the lines of these $k$ is generally upward. Notice
that $\rho_2$ reaches a maximum and then decreases to a constant
value, as can be seen in Fig. \ref{fig1}(c). By the end of the
initial stage at time $\tau=10$, the creation of smaller clusters,
with $2\leq k\leq 5$, has slowed down greatly relative to the
initial spurt for times $0<\tau<2$. Furthermore, the number of
clusters with more than 5 monomers is negligible. At $\tau=10$,
$\langle k\rangle \approx 2.69$, much smaller than the
equilibrium value $\langle k\rangle \approx \sqrt{\rho e^{\alpha
}}= \epsilon^{ -{1\over 2}} \approx 46.9$. To determine the time 
scales appropriate for exploring the subsequent kinetics, it is
highly instructive to plot the average cluster size $\langle
k\rangle$ as a function of time, based on the numerical solution.
Fig. \ref{fig4} is a log-log plot of $\langle k\rangle/e$ as a
function of $\tau$. It reveals an initial rapid growth of
$\langle k\rangle$ to a ``plateau value'' close to $e$, roughly
located in the interval $10<\tau < 100$. In the subsequent growth
after the plateau, large clusters with $k\gg 1$ eventually
appear. Fig. \ref{fig4} indicates that by time $\tau = 5\times
10^4$, $k$ clusters having $\langle k\rangle\approx 10$ are
prevalent.

Fig. \ref{fig2} shows frames at times $\tau=20$, $10^4$ and
$2\times 10^4$, thereby continuing those in Fig. \ref{fig1}. The
heavy dots correpond to $\tau=20$, which is well inside the
plateau phase. The histograms at $\tau=10^4$ and $2\times 10^4$
indicate the clear emergence of a continuum limit of the
kinetics.

In the time interval $2\times 10^4 < \tau < 5\times 10^5$, the
log-log plot of $\langle k\rangle/e$ as a function of $\tau$ in
Fig. \ref{fig4} is close to a straight line of slope $1/2$. This
strongly supports the existence of a self-similar stage of the
kinetics. The line graphs in Fig. \ref{fig5} depict $\langle k
\rangle^2\, r_k$ as a function of $x\equiv k/\langle k\rangle
$ for the times $\tau= 0.5 \times 10^5, 10^5, 1.5\times 10^5$.
They are nearly superimposed on top of each other. The heavy
dots correspond to the plateau time $\tau= 20$, so the change in
the distribution shape over the whole time span $20<\tau < 1.5
\times 10^5$ is not very great.

The self-similar stage is not the final chapter of the kinetics
story either. By $\tau = 10^6$, the linear dependence of
$\ln(\langle k\rangle/e)$ with ln$\tau$ breaks down. In fact, at
$\tau = 10^6$, $\langle k\rangle\approx 31.1$, which is
comparable to the equilibrium value of 46.9 mentioned before.
Evidently, there is a final stage of kinetics in which the size
distribution asymptotes to its equilibrium form. Fig. \ref{fig3}
is the final era of cluster aggregation, continued from
Fig. \ref{fig2}, in which snapshots of the size distribution are
taken at $\tau$ increments of $0.2\times 10^6$, from $0.2\times
10^6$ to $4\times 10^6$. Convergence to an exponential
distribution with $\langle k\rangle$ equal to the equilibrium
value of 46.9 is clear.

\setcounter{equation}{0}
\setcounter{figure}{0}
\section{Asymptotic theory of micellization}
\label{sec:asymptotics}
In this section, we shall interpret the numerical results shown
in Section \ref{sec:numerics} by using singular perturbation
methods; see Ref. \onlinecite{kevorkian} for a general
description thereof. 

\subsection{Initial transient}
Initially, $r_1(0) = 1$ and there are no multiparticle
aggregates. As we have seen in Section \ref{sec:numerics}, the
numerical solution of the complete model shows that there is an
initial transient stage during which dimers, trimers, etc.\ form
at the expense of the monomers, and that $r_k\approx 0$ for
sufficiently large $k$. Taking the $\epsilon\to 0$ limit of
Eqs.\ (\ref{bd10}) and (\ref{bd11}) yields the following planar
dynamical system:
\begin{eqnarray} 
{dr_{1}\over ds} & =& - (r_1+ r_c) , \label{bd13}\\  
{dr_{c}\over ds} &=& - r_c , \label{bd14}\\ 
{ds\over d\tau}  &=& r_1 , \label{bd15}
\end{eqnarray}  
in the {\em adaptive time} $s=\int_0^\tau \rho_1 d\tau$. The
general solution of the linear system (\ref{bd13}) - (\ref{bd14})
is
\begin{eqnarray} r_1 = (a - bs)\, e^{-s},\quad r_c = b\, e^{-s}
,  \nonumber
\end{eqnarray}  where $a$ and $b$ are arbitrary constants. Our
initial condition yields $a=b=1$, so that
\begin{eqnarray} r_1 = (1 - s)\, e^{-s},\quad r_c = e^{-s} ,
\label{bd16}
\end{eqnarray}  and from Eq.\ (\ref{bd15}),
\begin{eqnarray}
\tau = \int_0^s {e^{s}\over 1 - s}\, ds . \label{bd17}
\end{eqnarray}  Clearly, $\tau\to \infty$ corresponds to $s\to
1-$. At $s=1$, Eq.\ (\ref{bd16}) yields $r_1=0$, $r_c=e^{-1}$,
which are the limiting values of the variables $r_1$ and $r_c$
at the end of the initial stage. Eq.\ (\ref{bd9}) with
$\epsilon=0$ becomes $d(r_k e^s)/ds = r_{k-1} e^s$, which can be
solved recursively to yield
\begin{eqnarray} r_k = \left({s^{k-1}\over (k-1)!}- {s^{k}\over
k!} \right)\, e^{-s} .  \label{bd18}
\end{eqnarray}  As $\tau\to \infty$, $r_k \to (k-1) e^{-1}/k!$.
Since $r_6(1)= 0.00255$, after the initial transient stage there
are insignificant numbers of aggregates with more than 5
monomers. In fact, the average aggregate cluster size is $\langle
k\rangle = 1/r_c = e$, whereas at equilibrium, $\langle k
\rangle \sim \sqrt{\rho e^\alpha}\gg 1$. We therefore conclude
that there must be successive transients on time scales much
larger than $t= O(\epsilon)$.

\subsection{Intermediate transient}
Examination of the exact equation (\ref{bd9}) shows that when
$r_1$ decreases to size $O(\epsilon)$ but $r_2$, $r_3$, \ldots
are of order 1, all terms in its right hand side are
$O(\epsilon)$. This suggests rescaling $r_1 = \epsilon R_1$, so
that $\rho_1 = e^{-\alpha} R_1$, and using the original time
$t=\epsilon \tau$. Eq.\ (\ref{bd9}) becomes
\begin{eqnarray}
{dr_{2}\over dt} &=& - (R_1-1) (r_2-\epsilon
R_1)+ r_3-2 r_2 +
\epsilon R_1, \label{bd19}\\ {dr_k\over dt} &=& - (R_1-1) (r_k -
r_{k-1}) + r_{k+1} -2r_k + r_{k-1} ,\quad k\geq	2. \label{bd20}
\end{eqnarray}
The global identities (\ref{bd10}) and (\ref{bd11}) become
\begin{eqnarray}
(R_1 - 1)\, r_c - r_2 + \epsilon\, \left(
{dR_{1}\over dt} + R_1^2 + R_1 \right) = 0, \label{bd21}\\
{dr_c\over dt} + (R_1-1)\, r_c + \epsilon R_1 = 0, \label{bd22}
\end{eqnarray}
where now $r_c = \epsilon R_1 +\sum_{k=2}^\infty r_k \sim
\sum_{k=2}^\infty r_k$, as $\epsilon \to 0$. In the limit
$\epsilon\to 0$, $R_1 - 1 = r_2/r_c$ and Eq.\ (\ref{bd20})
becomes
\begin{eqnarray} {dr_k\over dt} &=& - {r_2\, (r_k -
r_{k-1})\over r_c} + r_{k+1} -2r_k + r_{k-1} ,\quad k\geq	2.
\label{bd23}
\end{eqnarray}
This is a closed system of equations for $r_2$, $r_3$, \ldots, to
be solved with the asymptotic values $r_k= (k-1)e^{-1}/k!$ as
initial conditions. It can be shown that the reduced versions of
Eqs.\ (\ref{bd22}), $\dot{r}_c = - (R_1-1) r_c$, and the
conservation condition, $\sum_{k=2}^\infty k r_k = 1$, are
upheld automatically by the solution of Eqs.\ (\ref{bd23}), so
that they are redundant for this stage.

The numerical solution of the reduced system of equations
(\ref{bd23}) for $r_k$, $k\geq 2$ closely approximates that of
the full system of kinetic equations at this stage. It can be seen
that more and more $r_k$ become different from zero as $t$
increases and that $r_k-r_{k-1}$ becomes small. This strongly
suggests that $r_k$ can be approximated by a continuum limit for
long times. To find the continuum limit, we set
\begin{eqnarray} 
r_k(t) \sim \delta^a\, r(x,T),\quad x = \delta\, k, \, T =
\delta^b\, t.   \label{bd24}
\end{eqnarray}
Here $\delta\to 0$ fixes the scale of $k= O(1/\delta)$, so that
$x$ is fixed at some value of order 1. $a$ and $b$ are positive
exponents to be determined. To find $a$, we use the conservation
condition, $\sum_{k=2}^\infty k r_k = 1$:
$$1 = \delta^{a-2}\, \sum_{k=2}^\infty (k\delta)\, r(k\delta,T)\,
\delta \sim \int_0^\infty x\, r(x,T)\, dx,
$$ 
provided $a=2$. The limiting form of the particle conservation
is thus
\begin{eqnarray}
 \int_0^\infty x\, r(x,T)\, dx = 1.   \label{bd25}
\end{eqnarray}
A similar calculation for the total number of clusters yields
$r_c \sim \delta\, \int_0^\infty r(x,T)\, dx$, which suggests the
definition
\begin{eqnarray} 
r_c \sim \delta\, R_c, \quad R_c \equiv
\int_0^\infty r(x,T)\, dx .
\label{bd26}
\end{eqnarray}  
We now substitute Eq.\ (\ref{bd24}) in Eq.\ (\ref{bd23}) and use
(\ref{bd26}) instead of $r_c$. The result is
\begin{eqnarray}
\delta^b\, {\partial r\over\partial T} \sim - {\delta^2\,
r(2\delta,T)\, [r(x,T) - r(x-\delta,T)])\over \delta\, R_c} +
r(x+\delta,T) -2r(x,T) + r(x-\delta,T) . \nonumber
\end{eqnarray}
The right hand side of this expression is of order $O(\delta^2)$,
so that the following distinguished limit is obtained if we set
$b=2$ and take $\delta\to 0$:
\begin{eqnarray} 
{\partial r(x,T)\over\partial T} = -{r(0,T)\over R_c(T)}\,
{\partial r(x,T)\over \partial x} + {\partial^2 r(x,T)\over
\partial x^2}. \label{bd27}
\end{eqnarray}  
For $k=2$, Eq.\ (\ref{bd23}) and the scaling (\ref{bd24}) with
$a=b= 2$ imply that $r(0,T)=0$. Therefore (\ref{bd27}) becomes the
simple diffusion equation:
\begin{eqnarray}
{\partial r\over\partial T} = {\partial^2 r\over \partial x^2},
\label{bd28}
\end{eqnarray}
for $x>0$, $t>0$ to be solved with the boundary condition $r(0,T)
=0$. 

The numerical solution of the discrete equations (\ref{bd23})
show that large aggregates do not emerge until $t\gg 1$. This
suggests that the appropriate solution of Eq.\ (\ref{bd28})
should be concentrated about $x=0$ as $T\to 0+$. That solution is
proportional to the $x$ derivative of the diffusion kernel,
\begin{eqnarray}
r(x,T) = -{\partial\over \partial x}\left(
{e^{-{x^2\over 4T}}\over \sqrt{\pi T}}\right) = {x\over
2\sqrt{\pi} T^{{3\over 2}}}\, \exp\left(-{x^2\over 4T}\right) .
\label{bd29}
\end{eqnarray}
The numerical prefactor is chosen so that particle conservation,
given by Eq.\ (\ref{bd25}), holds. It follows from Eq.\
(\ref{bd26}) that $R_c = (\pi T)^{-{1\over 2}}$. Hence the
average aggregate size is
\begin{eqnarray}
\langle k\rangle = {1\over \delta\, R_c} = {\sqrt{\pi T}\over
\delta} .  \label{bd30}
\end{eqnarray}  
In terms of the original variables $k$, $t$ and
$r_k$, the previous expressions are
\begin{eqnarray} 
r_k(t) \sim {k\over 2\sqrt{\pi}\, t^{{3\over 2}}}\,
\exp\left(-{k^2\over 4t}\right) ,  \label{bd31}\\
\langle k \rangle \sim \sqrt{\pi t}, \label{bd32}
\end{eqnarray}
as $t\to\infty$. These two equations yield
\begin{eqnarray}
\langle k\rangle^2\, r_k \sim {\pi k\over 2\langle
k\rangle}\,\exp\left[-{\pi\over 4}\, \left({k
\over\langle k\rangle }\right)^2\right]\,,
\label{bd32a} 
\end{eqnarray}
which resembles the behavior of the numerical solution of the
full kinetic model as indicated in Fig. \ref{fig5}. Notice that
the average cluster size $\langle k\rangle$ corresponding to the
solution of Eqs.\ (\ref{bd23}) (dotted line in Fig.
\ref{fig4}) approaches the value (\ref{bd32}) (straight line of
slope 1/2 in Fig. \ref{fig4}).

\subsection{Equilibrium transient} 
The large time limit of Eq.\ (\ref{bd31}) does not match the
equilibrium size distribution, which is $r_k \sim \epsilon
e^{-k\sqrt{\epsilon}}$ in the same scaled units; see Section
\ref{sec:model}. Thus the limit given by Eq.\ (\ref{bd31}) is
expected to break down when it predicts an average $\langle k
\rangle$ of the order of the equilibrium length $1/\sqrt{\epsilon
}$. According to Eq.\ (\ref{bd32}), this occurs at a time $\sqrt{
t} = O(\epsilon^{-{1\over 2}})$, i.e., $t = O(\epsilon^{-1})$. In
this third and final transient towards equilibrium, we set:
\begin{eqnarray} 
r_k(t) = \epsilon r(x,t), \quad x= \sqrt{\epsilon}\, k,\, T =
\epsilon\, t. \label{bd33}
\end{eqnarray}  
This is the same scaling as in Eq.\ (\ref{bd24}) with $a=b=2$ and
$\delta = \sqrt{\epsilon}$, and therefore we use here the same
notation for the variables. With this scaling, the scaled
particle conservation is
\begin{eqnarray*} 1 = \sum_{k=1}^\infty k\, r_k =
\epsilon^{{1\over 2}}\, \sum_{k=1}^\infty \epsilon^{{1\over
2}}\, k\, r(x,T),
\end{eqnarray*}  
and the limit $\epsilon\to 0$ yields
\begin{eqnarray}
 \int_0^\infty x\, r(x,T)\, dx = 1 . \label{bd34}
\end{eqnarray}  Similarly,
\begin{eqnarray} r_c\sim \epsilon^{{1\over 2}}\, \int_0^\infty
r(x,T)\, dx \equiv
\epsilon^{{1\over 2}}\, R_c . \label{bd35}
\end{eqnarray}  The scaled version of the global identity
(\ref{bd7}) is
\begin{eqnarray} R_c\, (R_1 -1)+ \epsilon^{{1\over 2}}\, R_1 +
\epsilon\, {dR_c
\over dT} = 0 .  \label{bd36}
\end{eqnarray}  Here $r_1 = \epsilon R_1 = \epsilon\,
r(\epsilon^{{1\over 2}},T)$. It follows from Eq.\ (\ref{bd36})
that
\begin{eqnarray} R_1 -1 = - {\epsilon^{{1\over 2}}\over R_c}  +
O(\epsilon) .
\label{bd37}
\end{eqnarray}  
The scaled kinetic equation (\ref{bd9}) is
\begin{eqnarray*}
\epsilon^3 {\partial r\over\partial T} = - \epsilon^2 (R_1 -
1)[r(x,T) - r(x-\epsilon^{{1\over 2}},T)] + \epsilon^2
[r(x+\epsilon^{{1\over 2}},T) -2 r(x,T) + r(x-\epsilon^{{1\over
2}},T)] .
\end{eqnarray*}
We now substitute Eq.\ (\ref{bd37}) in this
expression, divide it by $\epsilon^3$ and take the limit
$\epsilon\to 0$. The result is
\begin{eqnarray} {\partial r\over\partial T} = {1\over R_c(T)}\,
{\partial r\over\partial x} + {\partial^2 r\over\partial x^2}.
\label{bd38}
\end{eqnarray}
In these units, the average aggregate length is $\langle
x\rangle = 1/R_c$, and Eq.\ (\ref{bd38}) can be rewritten as
\begin{eqnarray}
{\partial r\over\partial T} = \langle x\rangle\, {\partial r\over
\partial x} + {\partial^2 r\over\partial x^2}, \label{bd39}
\end{eqnarray}
to be solved with the boundary condition
\begin{eqnarray}
r(0,T) = 1 ,  \label{bd40}
\end{eqnarray}
which follows from Eq.\ (\ref{bd37}) with $\epsilon\to 0$. It
can be straightforwardly checked that ${d\over dT}\int_0^\infty
x\, r(x,T)\, dx = 0$, and therefore $\int_0^\infty x\, r(x,T)\,
dx = 1$, provided $r(x,0)$ satisfies this particle conservation
condition.

We now have to show two things:
\begin{enumerate}
\item As $T\to 0+$, the solution of Eqs.\ (\ref{bd39}) and
(\ref{bd40}) is asymptotic \cite{kevorkian} to the right hand
side of Eq.\ (\ref{bd29}), the self-similar limiting solution of
the intermediate transient stage.
\item The solution of Eqs.\ (\ref{bd39}) and (\ref{bd40}) tends
to the equilibrium size distribution as $T\to \infty$.
\end{enumerate} Then the size distribution of the equilibration
transient as $T\to 0+$ matches the long time limit of the
previous intermediate stage, and tends towards equilibrium as
$T\to \infty$. This completes the description of the dynamics of
the aggregate size distribution.

\subsubsection{Matching with the intermediate transient stage}
We represent $r(x,T)$ as
\begin{eqnarray} r(x,T) = {1\over T}\, h(\zeta,T), \quad \zeta=
{x\over\sqrt{T}} .
\label{bd41}
\end{eqnarray}  
With prefactor $1/T$, the particle conservation equation
(\ref{e6}) and the total cluster density adopt the invariant forms
\begin{eqnarray}
\int_0^\infty \zeta\, h(\zeta,T)\, d\zeta =
1,\quad\quad\quad\quad
\quad\quad\quad\quad\quad\quad\quad\quad\quad\quad\label{bd42}\\
R_c(T) = \int_0^\infty r(x,T)\, dx = {1\over\sqrt{T}}\,
\int_0^\infty h(\zeta,T)\, d\zeta \equiv {h_c(T)\over\sqrt{T}}.
\label{bd43}
\end{eqnarray}  
Then 
\begin{eqnarray}
\langle x\rangle = {\sqrt{T}\over h_c(T)} .  \label{bd44}
\end{eqnarray}  Inserting this equation together with Eq.\
(\ref{bd41}) in Eq.\ (\ref{bd39}), we obtain
\begin{eqnarray} {\partial^2 h\over\partial\zeta^2} + h +
{1\over 2}\zeta {\partial h\over\partial \zeta} =
T\,\left({\partial h\over\partial T} + {\zeta h\over
h_c}\right), \label{bd45}
\end{eqnarray}  to be solved with the boundary condition
indicated by Eqs.\ (\ref{bd40}) and (\ref{bd41}):
\begin{eqnarray} h(0,T) = T .  \label{bd46}
\end{eqnarray}  
Asymptotic similarity as $T\to 0$ means that $h(\zeta,T)$ in Eq.\
(\ref{bd41}) has a limit $H(\zeta)$ as $T\to 0$. The limit
equations obtained from Eqs.\ (\ref{bd42}), (\ref{bd45}) and
(\ref{bd46}) are
\begin{eqnarray*} 
{\partial^2 H\over\partial\zeta^2} + H +
{1\over 2}\zeta {\partial H\over\partial \zeta} =
0\quad\mbox{in}\quad \zeta>0,\\  H(0) = 0, \\
\int_0^\infty \zeta\, H(\zeta)\, d\zeta = 1.
\end{eqnarray*}  
The unique solution of these equations is $H(\zeta)= \zeta
e^{-\zeta^2/4}/(2\sqrt{\pi})$, which is the right hand side of
Eq.\ (\ref{bd29}).

\subsubsection{Trend towards equilibrium}
The stationary solution of Eq.\ (\ref{bd39}) with the condition
(\ref{bd40}) is $r_e= e^{-x\langle x\rangle}$, and the particle
conservation condition gives $\langle x\rangle^2 =1$, so that
$\langle x\rangle =1$. Then the stationary solution of Eq.\
(\ref{bd39}) is $r_e= e^{-x}$, which is the sought equilibrium
solution. To show that $r(x,T)\to r_e(x)$ as $T\to \infty$, we
define the associated free energy
\begin{eqnarray}
f[r] = \int_0^\infty \left[-r + r\,\ln
\left({r\over r_0}\right)\right]\, dx - 1,  \label{bd47}\\
r_0 = e^{-x} , \label{bd48}
\end{eqnarray}
and show that it is a Lyapunov functional for Eq.\ (\ref{bd39}).
Notice that $\int_0^\infty r\, \ln r_0\, dx = - \int_0^\infty x\,
r\, dx=- 1$, and therefore $f[r]$ is the usual free energy, $f[r]
= \int_0^\infty (r\ln r - r)\, dx$.

Firstly, the standard inequality $x\ln x \geq x-1$ for positive
$x=r/r_0$ yields $f\geq - \int_0^\infty e^{-x}\, dx-1 = -2$, and
therefore $f$ is bounded below. Notice that $f[r_0]=-2$ at
equilibrium.

Secondly, time differentiation of Eq.\ (\ref{bd48}) yields
\begin{eqnarray*}
{df\over dT} = \int_0^\infty {\partial r\over
\partial T}\,\ln \left({r\over r_0}\right)\, dx.
\end{eqnarray*}
If we now substitute Eq.\ (\ref{bd39}), integrate by parts, and
use $r(0,T)= r_0(0)=1$ and $\int_0^\infty r\, dx = 1/\langle
x\rangle$, we obtain
\begin{eqnarray}
{df\over dT} = \langle x\rangle - \int_0^\infty
{1\over r}\,\left(  {\partial r\over \partial x}\right)^2\, dx
=\langle x\rangle
\left[1-\int_0^\infty r\, dx\, \int_0^\infty {1\over r}\,\left(
{\partial r\over \partial x}\right)^2\, dx\right].  \label{bd49}
\end{eqnarray}
The right hand side of this equation is less or equal than zero
because of the Cauchy-Schwarz inequality:
\begin{eqnarray*}
1 = r(0,T)^2 = \left(\int_0^\infty {\partial
r\over \partial x}\, dx\right)^2 \leq \left(\int_0^\infty
\left|{\partial r\over
\partial x}\right|\, dx\right)^2 \leq \int_0^\infty r\, dx\,
\int_0^\infty {1\over r}\,\left( {\partial r\over \partial
x}\right)^2\, dx.
\end{eqnarray*}
Therefore, we have proven that the free energy is a
Lyapunov functional. We can rewrite Eq.\ (\ref{bd49}) in
an equivalent form by defining $\tilde{r}_0=\exp[-x\langle x
\rangle]$, and using the identities:
\begin{eqnarray*}
\langle x\rangle = \langle x\rangle^2\int_0^\infty r\, dx =
\int_0^\infty r\, \left({\partial\ln \tilde{r}_0\over \partial
x}\right)^2\, dx,\\
\langle x\rangle = - \langle x\rangle\,\int_0^\infty {\partial r
\over\partial x}\, dx = \int_0^\infty r\, {\partial \ln r
\over\partial x}\,{\partial \ln \tilde{r}_0\over\partial x}\, dx,
\end{eqnarray*}
to obtain
\begin{eqnarray}
{df\over dT} = -\int_0^\infty r\, \left[ {\partial\over \partial
x}\,\ln\left({r\over \tilde{r}_0}\right)\right]^2\, dx \leq 0.
\label{bd50}
\end{eqnarray}
This equation shows that $r\to \tilde{r}_0$ as $T\to \infty$.
The particle conservation condition, $\int_0^\infty x
\tilde{r}_0 dx =1$ yields $\langle x\rangle^2 =1$, and therefore
$\tilde{r}_0=e^{-x}$.

\subsubsection{Approximation of the size distribution function by
matched asymptotic expansions }
\begin{figure}
\begin{center}
\includegraphics[width=7cm]{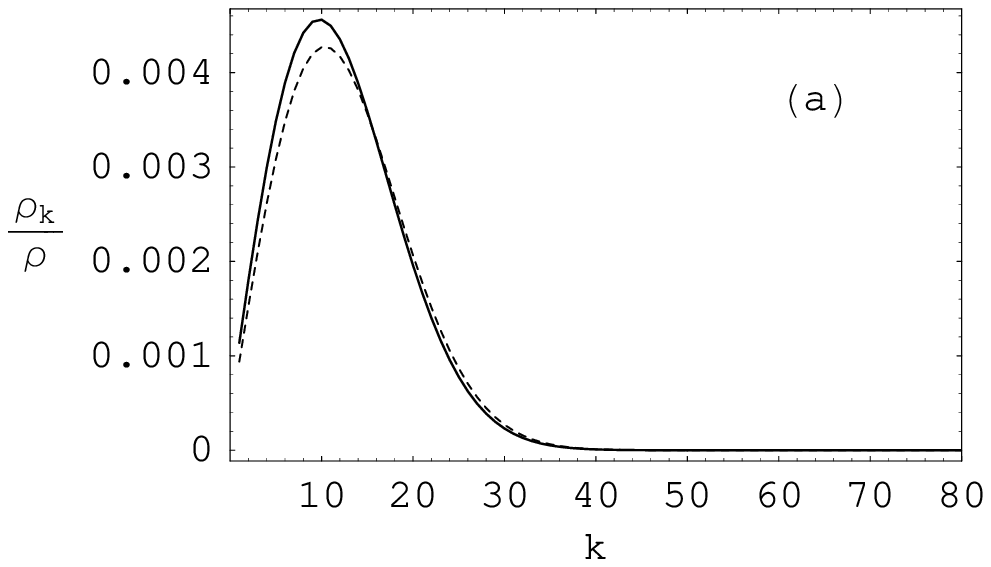}
\includegraphics[width=7cm]{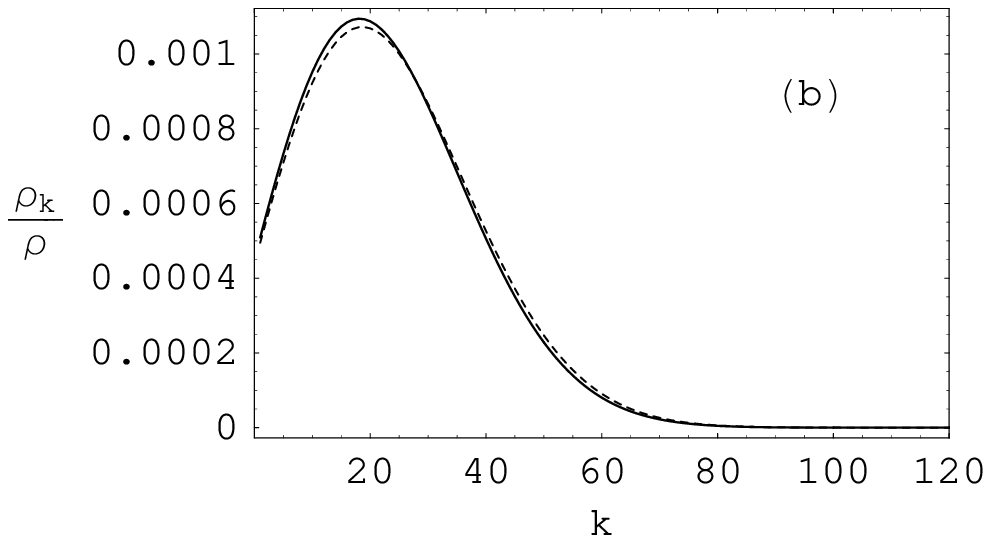}
\includegraphics[width=7cm]{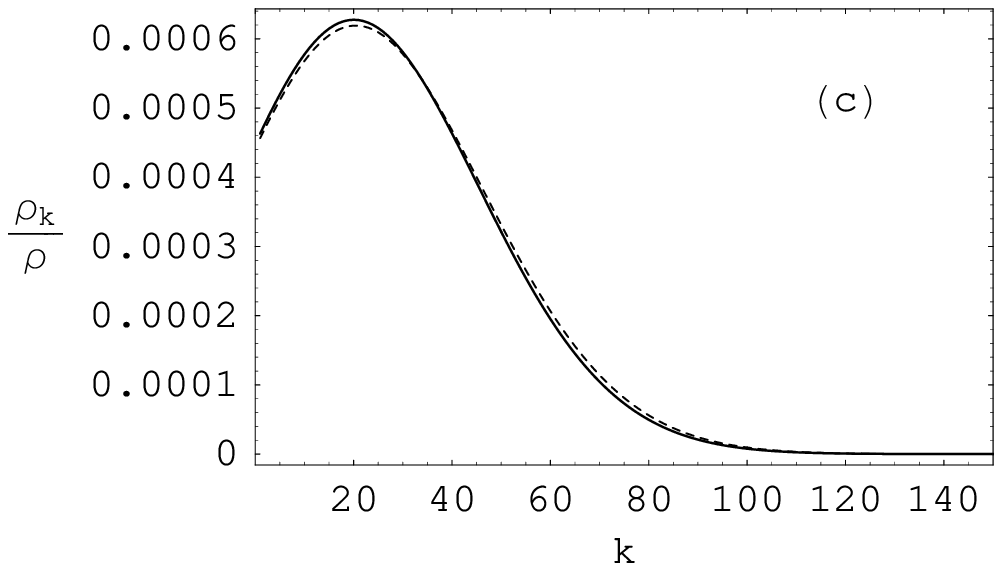}
\includegraphics[width=7cm]{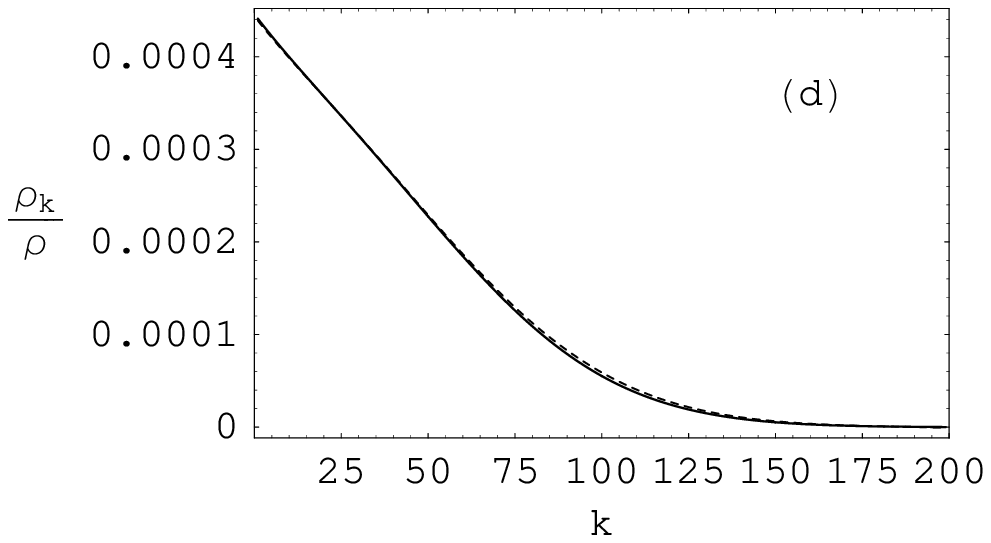}
\vspace{0.5 cm}
\caption{Comparison of the approximation (\ref{bd51}) (dashed
line) to the numerical solution of the full kinetic model (solid
line) for four different times $\tau$: (a) 100,000, (b) 500,000,
(c) 1,000,000, and (d) 3,000,000. Notice that the agreement
improves as the equilibrium distribution is approached.}
\label{fig6}
\end{center}
\end{figure}

An uniformly valid approximation to the size distribution
function can be easily formed from: (i) $r_k^{(1)}(\tau)$, given
by Eqs.\ (\ref{bd17}) and (\ref{bd18}), (ii) $r_k^{(2)}(t)$, which
solves the approximate system of Equations (\ref{bd23}) and $r_c =
\sum_{k=2}^\infty r_k$ with the initial conditions $r_k(0)= (k-1)
e^{-1}/k!$, and (iii) $r(x,T)$, which solves the nonlinear
Fokker-Planck equation (\ref{bd39}) with the condition
(\ref{bd40}) and it matches (\ref{bd29}) as $T\to 0+$. The result
is
\begin{eqnarray}
r_k^{(unif)}(\tau) = r_k^{(1)}(\tau) + r_k^{(2)}(\epsilon\tau)
+ \epsilon\, r(\sqrt{\epsilon} k,\epsilon^2\tau) - {k-1\over k!\,
e} -{k\over 2\sqrt{\pi}\, (\epsilon\tau)^{{3\over 2}}}\,
\exp\left(-{k^2\over 4\epsilon\tau}\right) . \label{bd51}
\end{eqnarray}
Figures \ref{fig6} compare the distribution function given by
Eq.\ (\ref{bd51}) to the numerical solution of the complete model
equations in times corresponding to the end of the intermediate
stage and the beginning of the equilibration stage. At these
times, $r_k^{(1)}=(k-1)/(k!\, e)$. We observe a good agreement
between approximate and numerical solutions, which improves as
the time elapses and the equilibrium distribution is approached. 
 
\setcounter{equation}{0}
\setcounter{figure}{0}
\section{Conclusions}
\label{sec:conclusion}
On the basis of a simple kinetic model and starting from the
initial state of pure monomers, we have shown that the process of
micellization of rod-like aggregates at high CMC occurs in three
separated stages or eras. In the first era, many clusters of
small size are produced while the number of monomers decreases
sharply. During the second era, aggregates are increasing
steadily in size and their distribution approaches a self-similar
solution of the diffusion equation. Before the continuum limit
can be realized, the average size of the nucleii becomes
comparable to its equilibrium value, and a simple mean-field
Fokker-Planck equation describes the final era until the
equilibrium distribution is reached. A continuum size
distribution does not describe micellization until the third era
has started: during the first two eras the effects of
discreteness dominate the dynamics.

In order to validate our theory by an experiment, it would be
important to measure the average cluster size as a function of
time as in Fig. \ref{fig4}: the multiscale behavior is more
clearly seen in this figure. To determine the time scale, we need
a measure of the cluster diffusion coefficient, $d$, which was
set equal to 1 in Section \ref{sec:model}. A convenient relation
could be Eq.\ (\ref{bd32}), which in dimensional units is $\langle
k\rangle \approx \sqrt{d\pi t}$. In case the self-similar size
distribution is not reached during the intermediate phase,
another way to determine $d$ is to study the equilibration era
and compare the experimentally obtained size distribution with the
numerical solution of the model. At equilibrium, $\langle k
\rangle^2 \approx \rho e^{\alpha}$, and this relation determines
the dimensionless binding energy $\alpha$.

\acknowledgments We thank A.\ Carpio for helpful discussions on
numerical solutions of the model. This work was carried out
during a visit of J.\ Neu's to the Universidad Carlos III de
Madrid, whose support we acknolwledge. J.A. Ca\~nizo was supported
by a FPU predoctoral scholarship granted by the Spanish Ministery
of Education. The present work was financed by the Spanish MCyT
grant BFM2002-04127-C02-01, by the Third Regional Research
Program of the Autonomous Region of Madrid (Strategic Groups
Action), and by the European Union under grant RTN2-2001-00349.

\end{document}